# Experimental research on the feature of an X-ray Talbot-Lau interferometer vs. tube accelerating voltage[*]

WANG Sheng-Hao (王圣浩)[1], Margie P. Olbinado[2], Atsushi Momose[2], HAN Hua-Jie (韩华杰)[1], HU Ren-Fang (胡仁芳)[1], WANG Zhi-Li (王志立)[1], GAO Kun (高 昆)[1], ZHANG Kai (张 凯)[3], ZHU Pei-Ping (朱佩平)[3], WU Zi-Yu (吴自玉)[1,3,†]

[1]National Synchrotron Radiation Laboratory, University of Science and Technology of China, Hefei 230027, China

[2]Institute of Multidisciplinary Research for Advanced Materials, Tohoku University, 2-1-1Katahira, Aoba-ku, Sendai, Miyagi 980-8577, Japan

[3]Institute of High Energy Physics, Chinese Academy of Sciences, Beijing 100049, China

**Abstract:** X-ray Talbot-Lau interferometer has been used most widely to perform X-ray phase-contrast imaging with a conventional low-brilliance X-ray source, it yields high-sensitivity phase and dark-field images of sample producing low absorption contrast, thus bearing tremendous potential for future clinical diagnosis. In this manuscript, while changing accelerating voltage of the X-ray tube from 35KV to 45KV, X-ray phase-contrast imaging of a test sample were performed at each integer KV position to investigate the characteristic of an X-ray Talbot-Lau interferometer (located in the Institute of Multidisciplinary Research for Advanced Materials, Tohoku University, Japan) vs. tube voltage. Experimental results and data analysis show that this X-ray Talbot-Lau interferometer is insensitive to the tube accelerating voltage within a certain range, fringe visibility around 44% is maintained in the aforementioned tube voltage range. This experimental research implies that potential new dual energy phase-contrast X-ray imaging strategy and rough refraction spectrum measurement is feasible with this X-ray Talbot-Lau interferometer.

**Key words:** X-ray Talbot-Lau interferometer, X-ray imaging, phase-contrast, tube accelerating voltage, X-ray tube

**PACS:** 87.59.-e, 07.60.Ly, 42.30.Rx, 87.57.-s

## 1. Introduction

X-ray phase-contrast imaging, which uses phase shift as the imaging signal, can provide remarkably improved contrast over conventional absorption-based imaging for weakly absorbing samples, such as biological soft tissues and fibre composites.[1-3] Over the last 50 years, several X-ray phase-contrast imaging methods has been put forward, they can be classified into crystal interferometer,[4-6] free-space propagation,[7,8] diffraction enhanced imaging,[9,10] and grating interferometer.[11,12] Although many excellent experiments research were accomplished based on these techniques, none of them has so far found wide applications in medical or industrial areas, where typically the use of a laboratory X-ray source and a large field of view are required. In 2006, Pfeiffer et al. [13] first developed and demonstrated a Talbot-Lau interferometer in the hard X-ray region with a conventional low-brilliance X-ray source, this can be considered as a great breakthrough in X-ray phase-contrast imaging, because it showed that phase-contrast X-ray imaging can be successfully and efficiently conducted with a conventional, low-brilliance X-ray source, thus overcoming the problems that impaired a wider use of phase-contrast in X-ray radiography and tomography, and many potential applications in biomedical imaging of this technique have been studied. [14-17]

[*] Supported by Major State Basic Research Development Program (2012CB825800), Science Fund for Creative Research Groups (11321503), and National Natural Science Foundation of China (11179004, 10979055, 11205189, 11205157).

[†] Corresponding author. E-mail: wuzy@ustc.edu.cn

One feature of X-ray Talbot-Lau interferometer is that theoretically speaking, with a giving phase grating, the ideal fringe visibility condition are valid only for a single wavelength (because phase shift and Talbot distance of the phase grating are directly related to the wave length of the X-ray). Typically, with an X-ray Talbot-Lau interferometer, accelerating voltage of the X-ray tube is chosen such that the mean energy of the radiation spectrum emitted from the X-ray tube matches the designed energy of the phase grating, and then the tube voltage is fixed in latter experiments. However, in practical experiments, for sample with different attenuation ability, tube voltage needs to be adjusted in order to get better image contrast. [18] Meanwhile, dual energy X-ray phase-contrast imaging is of interest to identify, discriminate and quantify materials between soft tissues, the present dual energy phase-contrast X-ray imaging strategy based on Talbot-Lau interferometer adopts two distinct energy spectra, whose mean energy separately match different orders of the Talbot distance, [19] one drawback of this method is that the fringe visibility at the high energy is too low (10%), so it is meaningful to explore the possibility of new dual energy strategy by choosing two energy spectrum around the designed energy of the phase grating for a certain Talbot distance. Upon the aforementioned two points, the feature of X-ray Talbot-Lau interferometer vs. tube accelerating voltage is of significances to be investigated.

In the following sections, we will demonstrate experimental research on the characteristic of an X-ray Talbot-Lau interferometer (located in the Institute of Multidisciplinary Research for Advanced Materials, Tohoku University, Japan) vs. tube accelerating voltage. Firstly, experimental setup, image acquisition manner and data post-processing would be introduced, and then X-ray imaging result of a test sample and the curve of fringe visibility vs. tube voltage are presented, at the same time, quantitative calculation of $\delta$ (decrement of the real part of the complex refractive index) and $\mu$ (linear attenuation coefficient) of the test sample by numerical fitting at each tube voltage position are demonstrated. After that, $\sigma$ (standard deviation) of absorption, refraction and visibility contrast image will be analyzed vs. tube voltage. Finally, we will discuss the experimental data and talk about some potential new usages of this X-ray Talbot-Lau interferometer.

## 2 Meterials and methods

### 2.1 Experimental setup and working principle

The grating-based X-ray phase-contrast imaging were carried out with an X-ray Talbot-Lau interferometer located at the Institute of Multidisciplinary Research for Advanced Materials, Tohoku University, Japan. Fig. 1 is the schematic of this X-ray interferometer. It is mainly made up of an X-ray tube, an X-ray detector and three micro-structured gratings, which are assembled on multi-dimensional motorized optical displacement tables. The X-ray is generated from a tungsten rotating anode X-ray source. The source grating G0 (period 22.7 μm, gold height 70 μm, size 20×20 mm$^2$) is positioned about 80 mm from the emission point inside the X-ray source, the beam splitter grating G1 (period 4.36 μm, Si height 2.43 μm, size 50×50 mm$^2$) which induced a phase shift of $\pi/2$ approximately at 27 keV, is placed 106.9 mm from G0 behind the gantry axis, and the sample is mounted closely before G1. The analyzer grating G2 (period 5.4 μm, gold height 65 μm, size 50×50 mm$^2$) is positioned about 10 mm before the detector, the distance between G1 and G2 is 25.6 mm. The X-ray detector is a combination of a scintillator and a CCD camera connected via a fiber coupling charge coupled device, its pixel size is 18×18 um$^2$ and the effective receiving area is 68.4×68.4 mm$^2$.

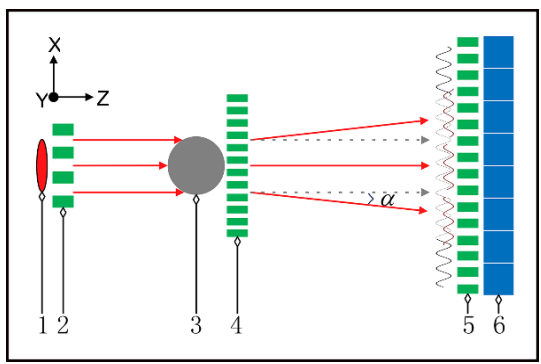

Fig. 1. (color online). Schematic of the X-ray Talbot-Lau interferometer. 1. X-ray source, 2. Source grating (G0), 3. Sample, 4. Beam splitter grating (G1), 5. Analyzer grating (G2), 6. X-ray detector.

As illustrated in Fig. 1, working principle of this Talbot-Lau interferometer is briefly described as follow, the source grating G0, an absorbing mask with transmitting slits, placed close to the X-ray tube anode, creates an array of line sources. Each line source is sufficiently small to prepare the required transverse coherence at the second grating, grating G1 is then coherently illuminated and self-image of G1 forms in the plane of grating G2 by the Talbot effect.[20] The images coming from all line sources look identical but they are shifted by multiples of the grating constant and therefore overlap in spite of their lack of mutual coherence. And then moire fringe is observed by the superposition of the self-image and G2 pattern in the plane of the X-ray detector. The differential phase-contrast image information process essentially relies on the fact that the sample placed in the X-ray beam path causes slight refraction of the beam transmitted through the object. And the fundamental idea of differential phase-contrast imaging depends on locally detecting these angular deviation, the angular $\alpha$ is proportional to the local gradient of the object's phase shift, and can be quantified as:

$$\alpha = \frac{\lambda}{2\pi}\frac{\partial \Phi(x,y)}{\partial x}. \qquad (1)$$

Where $\Phi(x,y)$ is the phase shift of the wave front, and $\lambda$ represents wavelength of the radiation. Determination of the refraction angle can be achieved by combining the moire fringe and phase-stepping technique,[21] a typical wave front measurement strategy, which contains a set of images taken at different positions of G2. When G2 is scanned along the transverse direction, the intensity signal of each pixel in the detector plane oscillates as a function of the grating position. By Fourier analysis, for each pixel, the shift curve of these oscillations, sample's conventional transmission, refraction and scattering signal defined by Pfeiffer et al.[22] can be simultaneously retrieved.[12,22]

2.2 Image acquisition and data post-processing

The sample we used is a PMMA cylinder, its diameter is 5mm and the length is about 100 mm. During the experiment, the X-ray generator was operated with tube current of 45 mA. After fine alignment of the three gratings and the sample stage, the tube voltage was tuned to 35KV, then 5 steps were adopted during the phase stepping scan, and for each step, 20 seconds was taken to capture a raw image. For the measurement of the background of the interferometer, a same phase stepping scan was performed after removing the PMMA cylinder. And then, we increased the tube voltage one KV by one KV to 45KV, at each position the similar experimental procedure was applied to carry out the phase-contrast X-ray imaging of the PMMA cylinder.

In the data post-processing, absorption image $A(m,n)$, refraction image $\alpha(m,n)$ and the normalized visibility contrast signal $V(m,n)$ of the sample were retrieved by a LabVIEW-based executive program using the NI Vision Development Module, the computing formula is as following:

$$A(m,n) = \ln\left[\frac{\sum_{k=1}^{N} I_k^s(m,n)}{\sum_{k=1}^{N} I_k^b(m,n)}\right], \quad (2)$$

$$\alpha(m,n) = \frac{p_2}{2\pi d} \times \arg\left[\frac{\sum_{k=1}^{N} I_k^s(m,n) \times \exp\left(2\pi i \frac{k}{N}\right)}{\sum_{k=1}^{N} I_k^b(m,n) \times \exp\left(2\pi i \frac{k}{N}\right)}\right], \quad (3)$$

$$V(m,n) = \frac{\sum_{k=1}^{N} I_k^b(m,n)}{\sum_{k=1}^{N} I_k^s(m,n)} \times \frac{\text{rem}\left[\sum_{k=1}^{N} I_k^s(m,n) \times \exp\left(2\pi i \frac{k}{N}\right)\right]}{\text{rem}\left[\sum_{k=1}^{N} I_k^b(m,n) \times \exp\left(2\pi i \frac{k}{N}\right)\right]}. \quad (4)$$

Here $N$ is the number of step during the phase stepping scan in one period of G2, $I_k^s(m,n)$ and $I_k^b(m,n)$ stand for the gray value of pixel $(m,n)$ at the $k^{th}$ step of the scan with and without sample, respectively. $p_2$ is the period of grating G2, and $d$ represents the distance between grating G1 and G2.

## 3 Experiment results and data analysis

### 3.1 X-ray imaging results and the curve of visibility vs. tube voltage

Fig. 2 shows X-ray imaging results of the PMMA cylinder, raw images were acquired while the X-ray tube was operated at tube accelerating voltage of 35KV. Fig. 2(a) stands for the conventional absorption image, while Fig. 2(b) is the differential phase contrast signal, and Fig. 2(c) depicts the normalized visibility contrast image. All the images are displayed on a linear gray scale and are windowed for optimized appearance. Here we want to point out that there is no visual differences between the X-ray imaging results obtained at 35KV tube accelerating voltage and those produced from other tube voltages.

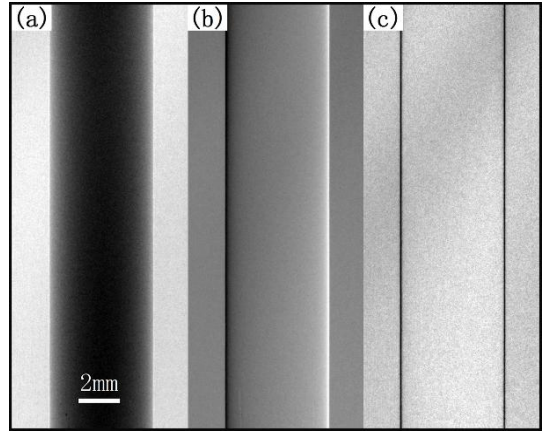

Fig. 2. X-ray imaging results of the PMMA cylinder, raw images were obtained at tube voltage of 35KV. (a) is the conventional absorption image, while (b) stands for the refraction signal, and (c) shows the normalized visibility image. All the images are displayed on a linear gray scale and are windowed for optimized appearance.

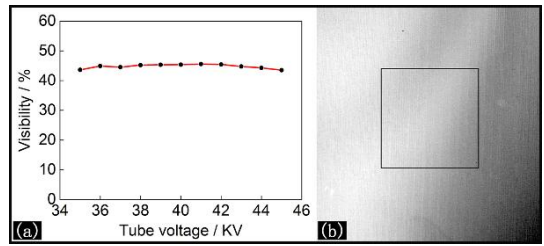

Fig. 3. (color online) (a) is the curve of fringe visibility vs. tube accelerating voltage, and (b) stands for the visibility image without sample at 35KV, the mean value in the black rectangular (800 pixels ×800 pixels) is regarded as the fringe visibility when plotting the curve in (a).

The fringe visibility of the X-ray Talbot-Lau interferometer at each adopted tube voltage position was obtained by computing the visibility image without sample in the beam path, and the curve of visibility vs. tube voltage is demonstrated in Fig. 3(a), while Fig. 3(b) depicts the visibility image of the Talbot-Lau interferometer at 35KV tube accelerating voltage. The mean value in the black rectangular (800 pixels×800 pixels) as shown in Fig. 3(b) was regarded as the visibility when plotting the curve in Fig. 3(a). Here it should be noticed that the vertical stripes in the visibility image as demonstrated in Fig. 3(b) well reveal some interesting details of the background of the Talbot-Lau interferometer, while Au grids of the grating in the beam path act as important mediums for visibility signal retrieval, both the still existing photoresists and the silicon wafers of the background can be regarded as the sample under analysis.

3.2 $\delta$, $\mu$ **of PMMA and** $V^s/V^b$ **vs. tube voltage**

of the complex refractive index of PMMA) vs. tube accelerating voltage is displayed in Fig. 4(a), while Fig. 4(b) and Fig. 4(c) shows the quantitative computing process of $\delta$ by numerical fitting at 35KV tube voltage. The black curve in Fig. 4(c) is cross section of the refraction signal chosen as shown in Fig. 4(b), here 30 rows were adopted to decrease the noise of the cross section, and the red line in Fig. 4(c) is the best fitting curve when $\delta$ equals 4.27399E-7, the fitting formula we used is:

$$\theta = \frac{2\delta x}{\sqrt{R^2 - x^2}} . \qquad (5)$$

Where $R$ represents radius of the PMMA cylinder, $x$ is the variable, and $\theta$ is the refraction angle, the fitting is performed based on the Levenberg-Marquardt algorithm.[23]

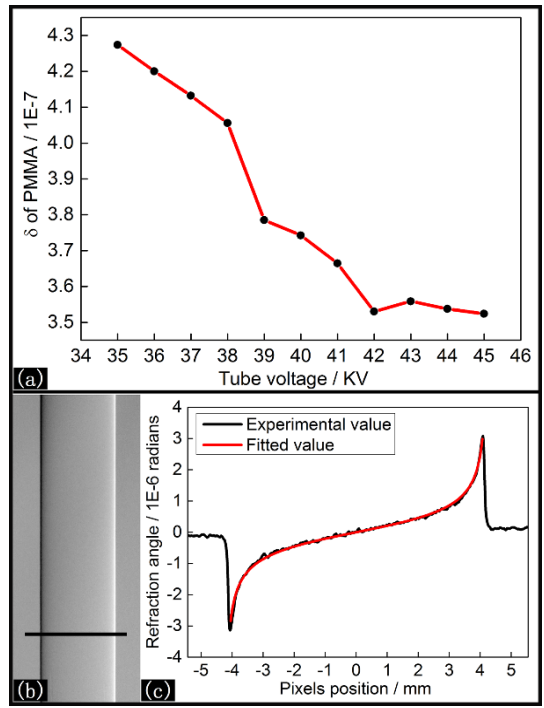

Fig. 4. (color online) (a) shows $\delta$ of PMMA vs. tube accelerating voltage, while (b) and (c) depict the computing process of $\delta$ at 35KV tube voltage. The black curve in (c) is cross section of the refraction signal chosen as shown in (b), and the red line in (c) is the best fitting curve when $\delta$ equals 4.27399E-7.

The curve of $\delta$ (decrement of the real part

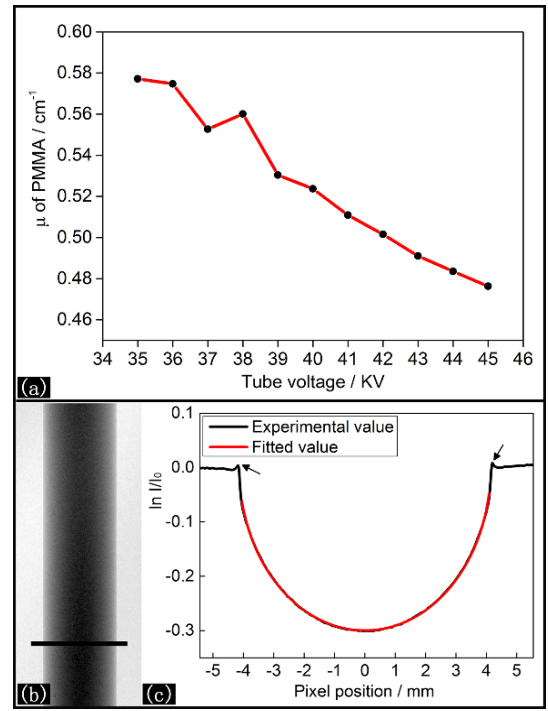

Fig. 5. (color online) (a) shows $\mu$ of PMMA vs. tube accelerating voltage, while (b) and (c) depict the computing process of $\mu$ at 35KV tube voltage. The black curve in (c) is section profile of the absorption image chosen as shown in (b), and the red line in (c) is the best fitting curve when $\mu$ equals 0.57724 cm$^{-1}$. The two small peaks as indicated by the black arrows in (c) show edge enhanced phenomenon in the retrieved absorption image.

Similarly, the curve of $\mu$ (linear attenuation coefficient of PMMA) vs. tube accelerating voltage is computed and demonstrated in Fig. 5, in contrast with the numerical fitting formula in generating $\delta$, the fitting equation we used here is:

$$\ln\frac{I}{I_0} = -\mu \times 2\sqrt{R^2 - x^2} . \qquad (6)$$

Where $I$ and $I_0$ depict gray value of the obtained X-ray image with sample and without sample respectively.

The visibility signal defined as $V^s/V^b$ (here $V^s$ and $V^b$ represent visibility image of the Talbot-Lau interferometer with and without sample, respectively) [22] at each adopted tube voltage position was calculated and the curve of $V^s/V^b$ vs. tube voltage is demonstrated in Fig 6.(a). And Fig 6.(b) shows image of $V^s/V^b$ yielded at tube voltage of 35KV, pixels of the black line located inside the rectangular (2 pixels ×10 pixels) were averaged to generate the curve.

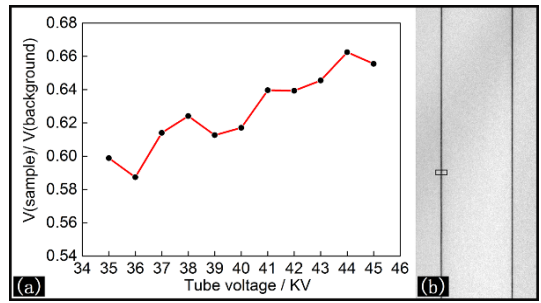

Fig. 6. (color online) (a) is the curve of $V^s/V^b$ vs. tube voltage, and (b) shows image of $V^s/V^b$ yielded at tube voltage of 35KV, pixels of the black line located inside the rectangular (2 pixels ×10 pixels) were averaged to generate the curve.

### 3.3 $\sigma$ of absorption, refraction and visibility signals vs. tube voltage

$\sigma$ (standard deviation) of absorption, refraction, and visibility image vs. tube accelerating voltage are demonstrated in Fig. 7(b), Fig. 7(d) and Fig. 7(e), respectively. Pixels in the black rectangular (30 pixels ×100 pixels) as shown in Fig. 7 (a), Fig. 7 (c) and Fig. 7 (e) were used to compute $\sigma$ of the three images respectively.

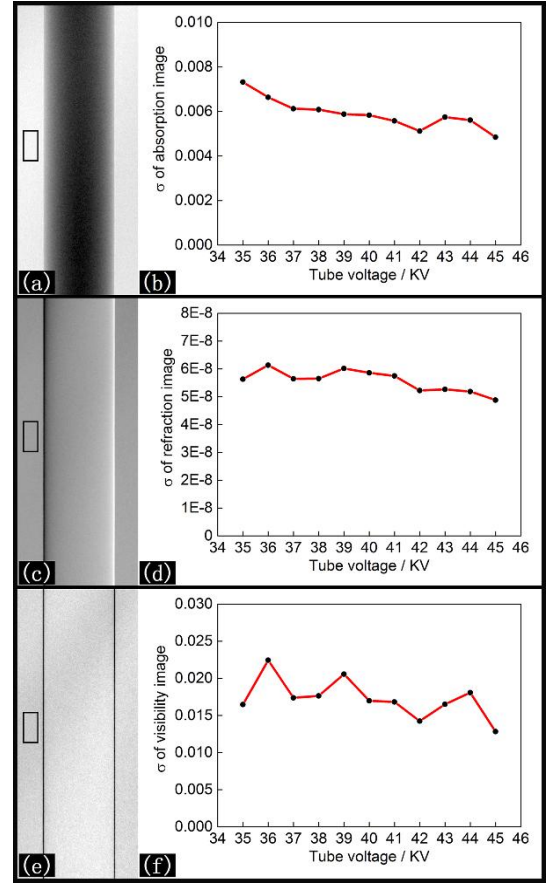

Fig. 7. (color online) (b), (d) and (e) depict respectively $\sigma$ of absorption, refraction, and visibility image vs. tube voltage. Pixels in the rectangular (30 pixels ×100 pixels) as shown in (a), (c) and (e) were used to compute $\sigma$ of the absorption/refraction/visibility image respectively.

### 4 Discussion

Fig. 2 represents three imaging signals of the PMMA cylinder, it should be pointed out that the retrieved absorption image as shown in Fig. 2(a) is a pseudo-absorption images, which is identical to the transmission image that would be obtained if the Talbot-Lau interferometer was not existing in the beam path. It contains the projected absorption coefficient and edge-enhancing Fresnel diffraction contrast. [24, 25] We can get the evidence in Fig. 2(a) that the border of the PMMA cylinder is slightly brighter than other part of the background, and also in Fig. 5(c) the cross section of the retrieved absorption image we can find a small peak appearing at both margin of the cylinder as indicated by the black arrows.

As shown in Fig. 3(a), it was observed that the fringe visibility is increasing slowly from 35KV to 41KV, and after that it turns declining. This phenomenon can be well explained by the point that the phase grating in the X-ray Talbot-Lau interferometer has a designed working energy (here it is 27 keV). The optimal fringe visibility is achieved when the mean energy of the emitted spectrum of the X-ray tube well matches the designed energy, and once the mean energy of the emitted spectrum deviates the designed working energy of the phase grating, worse visibility would be obtained. However, what should be pointed out at here is that fringe visibility of this X-ray Talbot-Lau interferometer changed very slightly and still maintain at a relatively very high lever in the tube voltage range of 35-45KV (the worst visibility we obtained in the experiment is 43% with 45KV tube voltage). Since fringe visibility is a key parameter of Talbot-Lau interferometer for refraction signal retrieval, [26-28] we think the aforementioned measured superior characteristic of this X-ray Talbot-Lau interferometer can be well utilized to perform experiments in some situations where the tube voltage are supposed to be changed, and also high fringe visibility is required. For example, situation (1): when investigating samples with different attenuation ability, the accelerating voltage of the X-ray tube needs to be adjusted in order to get optimal image contrast. [18] Situation (2): in dual energy X-ray phase-contrast imaging, a low energy spectra and a high one are adopted separately by changing the accelerating voltage of the X-ray tube. In the present dual energy X-ray phase contrast imaging strategy based on Talbot-Lau interferometer, the two energy spectra are chosen such that mean energy of the two spectra match different orders of the Talbot distance, [19] one drawback of this method is that the fringe visibility at the high energy is too low (10%). On the contrary, based on our obtained fringe visibility while changing tube voltage around the designed optima value, we think another new dual energy X-ray phase-contrast imaging strategy producing high image quality are available by choosing two energy spectra around the designed energy of the phase grating, even though the optional tube voltage range is limited in a certain extent. In our experiment with PMMA cylinder, dual energy X-ray imaging phase-contrast data was obtained that $\delta$ is 4.27E-7 (at 35KV), and 3.52E-7 (at 45KV), while $\mu$ is 0.57724 cm$^{-1}$ (at 35KV) and 0.47634 cm$^{-1}$ (at 45KV). Situation (3): in rough refraction spectrum measurement, as shown in Fig. 4(a) and Fig. 5(a), $\delta$ and $\mu$ of PMMA is measured with a conventional X-ray tube source in tube voltage range of 35-45KV, this measurement can find some potential applications in element recognition and material discrimination.

Also we find as shown in Fig. 6(a) that the visibility signal of the PMMA is related to the tube accelerating voltage, potential usage of this feature is similar to that of the other two signals.

Finally in the previous section, we calculated $\sigma$ (standard deviation) of the absorption, refraction and visibility contrast signals and plotted the curve of $\sigma$ vs. tube voltage as shown in Fig. 7. Theoretical speaking, $\sigma$ of the refraction signal retrieved from an X-ray Talbot-Lau interferometer is closely related to the fringe visibility, while the absorption and scattering signal are not. [29] However, as shown in Fig. 7(b), Fig. 7(d) and Fig. 7(f), it was observed that $\sigma$ of all the three signals is decreasing slowly with the increasing of the tube voltage, this phenomenon can be explained by the fact that the luminous flux emitted from the X-ray tube is roughly proportional to the second order of its tube voltage, and the effect of the slight change of the fringe visibility in the tube voltage range of 35-45KV imposed on the noise of the refraction image is negligible. We think the analysis of $\sigma$ in the refraction signal vs. tube voltage is a strong supporting evidence that this X-ray Talbot-Lau interferometer is insensitive to the tube voltage, so the aforementioned discussion by judging fringe visibility, that this X-ray Talbot-Lau interferometer can be well utilized to perform experiments in some situations where the tube voltage is supposed to be changed and also imaging quality should be guaranteed is valid.

## 5 Conclusion

In conclusion, characteristic of an X-ray Talbot-Lau interferometer vs. tube accelerating voltage was studied by investigating a test sample while changing tube accelerating voltage. Experimental results and data analysis show that this Talbot-Lau interferometer is insensitivity to the tube accelerating voltage from 35 KV to 45KV. This experiment research implies that potential new dual energy phase-contrast X-ray imaging strategy and rough refraction spectrum measurement within a certain range are feasible with this X-ray Talbot-Lau interferometer.

## Acknowledgements

The authors greatly acknowledge Murakami Gaku, Wataru Abe, Taiki Umemoto and Kosuke Kato (Institute of Multidisciplinary Research for Advanced Materials, Tohoku University, Japan) for their kind help while conducting the experiments. Also the authors want to acknowledge the financial funding support from Japan-Asia Youth Exchange program in Science (SAKURA Exchange Program in Science) administered by the Japan Science and Technology Agency.